\documentclass[11pt]{article}
\usepackage{moriond,epsfig}

\newcommand{\alps}{\alpha_s}
\def\be{\begin{equation}}
\def\ee{\end{equation}}
\def\bea{\begin{eqnarray}}
\def\eea{\end{eqnarray}}

\begin{document}
\vspace*{4cm}
\title{Jet physics and strong coupling at HERA}

\author{ M. Gouzevitch (on behalf of H1 and ZEUS collaborations) }

\address{The H1 Collaboration, DESY, Notkestrasse 85, 22607 Hamburg, Germany}

\maketitle\abstracts{
	Jet production in electron-proton scattering at HERA provides an
important testing ground for Quantum Chromodynamics (QCD). The inclusive jet and multi-jet cross sections recently measured by H1 and ZEUS collaborations allow a precise determination of the strong coupling and test of its running. Additionally, a measurement of the angular correlations in the 3-jet events gives a handle on the fundamental gauge structure of the QCD.}

\section{Jet production at HERA}
	
In \textit{ep} collisions at HERA one distinguishes two processes, according to the virtuality $Q^2$ of the exchanged boson, DIS and
photo-production. 

In DIS a highly virtual boson ($Q^2>1$~GeV$^2$) interacts
with a parton carrying a momentum fraction of the proton. The Born level contribution to DIS generates no transverse momentum in the Breit frame, where the virtual boson and the proton collide head on. Significant transverse momentum $P_T$ in the Breit frame is produced at leading order (LO) in the strong coupling $\alpha_s$ by the QCD-Compton and boson-gluon fusion processes. 

In direct photo-production the quasi-real photon ($Q^2<1$~GeV$^2$)
interacts with a parton from the proton. In resolved photo-production the photon 
behaves as a hadron and a parton from the photon, carrying a fraction
of its momentum, enters the hard scattering with the proton and gives rise to jet production.

In the analyses presented here jets are defined using the $k_T$ clustering
algorithm. The associated cross-sections are collinear and
infrared safe and therefore well suited for comparison with predictions from
fixed order QCD calculations.  For DIS, the jet algorithm is
applied in the Breit frame, and for photo-production in a photon-proton
collinear frame, the laboratory frame.

\section{Gauge Structure of QCD}

\begin{figure}
\centering
\psfig{figure=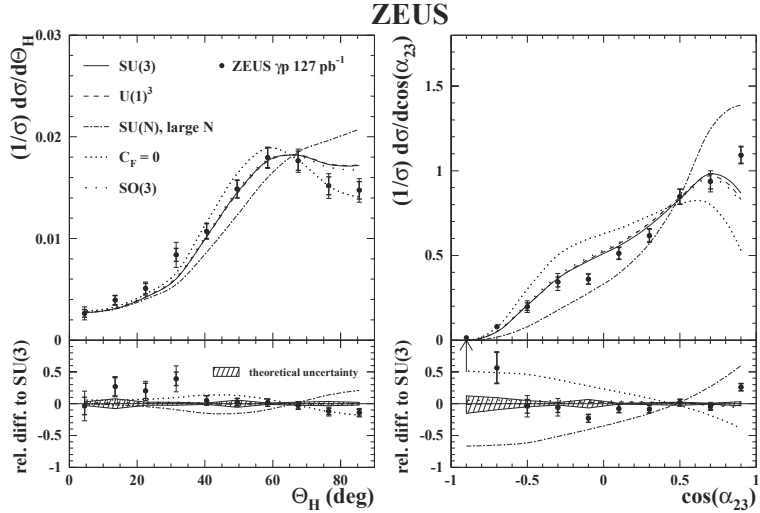,height=2.8in}
\caption{Differential 3-jet cross sections normalised to the shape in photoproduction as function of $\Theta_H$, the angle between the plane determined by the highest-transverse-energy jet and the beam and the plane determined by the two jets with lowest transverse energy (left) and the angle between second and third jet (right).
\label{fig:angcorr}}
\end{figure}

The angular correlations in the 3-jet cross sections were measured by the ZEUS collaboration in photoproduction and DIS based on a sample of about $130$~pb$^{-1}$ collected between 1995 and 2000~\cite{bib:Angular}. The transverse momentum $P_T$ of jets is required to exceed 14 GeV, if $Q^2<1~$GeV$^2$, or 
\mbox{8 GeV} for the first jet and 5 GeV for second and third jets, if $Q^2>125~$GeV$^2$. The differential cross sections were normalised in shape by the total 3-jet cross section in order to reduce the sensitivity to the scales, parton density functions (PDFs) and to the strong coupling constant and its running. In a given angular distribution, it may be possible to distinguish a
particular component of the hard scattering, e.g. its shape may distinguish between a two-boson-fermion vertex or a tri-boson vertex. The absolute contribution to the cross sections of each of those vertices is given by the colour factors which represents a signature of the underlying gauge group. An example of such the angular correlations is given in figure \ref{fig:angcorr}.

The theoretical uncertainties, typically of the order of $5\%$, are dominated by the experimental uncertainties of typically $10\%$. The main impact to the experimental uncertainty comes from the limited statistics (inner error bars) and from systematic uncertainties (outer error bars) dominated by the model dependence of data correction. This measurement rule out some of the choices of underlying gauge groups, like SU(N) with large N or $C_F=0$, but further improvements in sensitivity are needed to distinguish between SU(3), SO(3) and U(1)$^3$.

\section{Strong Coupling Determination}

\subsection{Jets cross sections}

In a ZEUS photoproduction analysis~\cite{bib:ZEUSPhP} the inclusive jet cross sections were measured by requiring the jet $P_T$ above
17~GeV and the jet pseudorapidity within \mbox{$-1.0<\eta^{\rm Lab}<2.5$}. The measured cross-sections are corrected for detector acceptance using leading order Monte Carlo event generators. The overall experimental systematic uncertainty of typically 10 to 15\% is dominated by the uncertainty on the absolute
energy scale of the hadronic calorimeters and the model dependence of data correction.

A jet measurement in DIS was recently performed by the H1 collaboration in two kinematic regimes. The low $Q^2$ data~\cite{bib:LowQ2}, corresponding to $5 < Q^2 < 100$~GeV$^2$,  are selected by requiring the scattered electron to be measured in the Spaghetti endcap Calorimeter. The high $Q^2$ data~\cite{bib:HighQ2}, corresponding to $150 < Q^2 < 15000$~GeV$^2$, are selected by requiring the scattered electron to be measured in the Liquid Argon barrel Calorimeter. At low $Q^2$ a sample of $44$~pb$^{-1}$ collected between 1999 and 2000 is used, whereas at high $Q^2$ the analysis is based on nearly the full H1 data sample of about $400$~pb$^{-1}$ collected between the years 1999 and 2007.

\begin{figure}

\centering
\psfig{figure=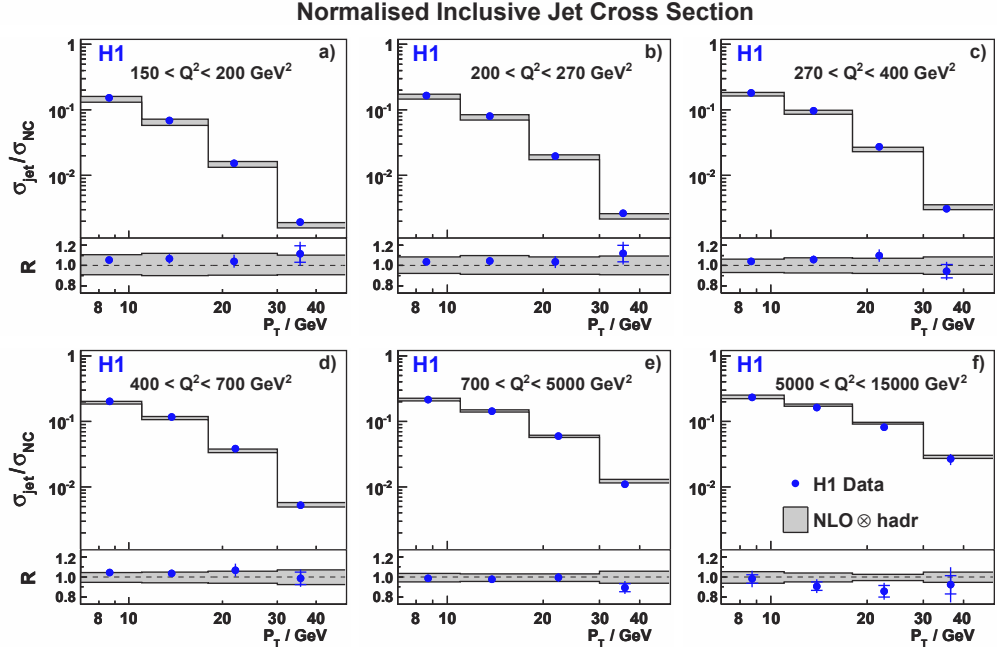,height=3.4 in}
\begin{flushleft}
\caption{The normalised inclusive jets cross sections in DIS as function of $P_T$ of jets in different $Q^2$ regions.
\label{fig:InclJets}}
\end{flushleft}
\end{figure}

 The inclusive jet cross sections were measured in low $Q^2$ regime by requesting \mbox{$P_T>5$~GeV} and $-1.0 < \eta^{\rm Lab} < 2.5$. At high $Q^2$ the cross sections was measured based on the inclusive jets with $7 < P_{T} < 50$~GeV and on 2-jet (3-jet) events containing at least 2 (3) jets with $5 < P_{T} < 50$~GeV. A more restrictive pseudorapidity cut was applied $-0.8 < \eta^{\rm Lab} < 2.0$ to ensure a good calibration of the jets. The jet cross sections at high $Q^2$ are normalised to the inclusive DIS cross sections in order to reduce the sensitivity to the normalisation uncertainties. The normalised jet cross sections as functions of $Q^2$ and $P_T$ are shown in figure \ref{fig:InclJets}. 
	
	One of the main sources of experimental uncertainties at low and high $Q^2$ remains the uncertainty on the absolute calibration of the hadronic energy scale with an impact on the cross sections of about 1 to 5\%. The detector correction factors show an uncertainty due to the MC model dependence which amounts typically to $1$ to 10$\%$.

\subsection{Determination of the strong coupling}

The strong coupling is extracted from the data by a minimal $\chi^2$ fit procedure where the value of the strong coupling at the $Z$ boson mass, $\alpha_s(M_Z)$, is taken to be the only free parameter of the theory. In ZEUS photoproduction analysis the experimental uncertainty is estimated by the \textsl{offset method} adding in quadrature the deviation of $\alps$ from the central value when the fit is repeated with independent variations of various experimental sources: 

\begin{eqnarray}
\nonumber
\alpha_s(M_Z) = 0.1223 ~\pm 0.0022 \,\mathrm{(exp.)} ~ ^{+0.0029}_{-0.0030}\,\mathrm{(th.)}\,. 
\end{eqnarray}
\noindent 

The theoretical uncertainty contains a dominating part coming from terms beyond NLO estimated using the band method of \textsl{Jones et al.} \cite{bib:Jones}, added in quadrature to the uncertainties on the hadronisation corrections and to the uncertainties on the proton and photon PDFs parameterisation. The total theoretical uncertainty amounts to $2.5\%$.

In the H1 analysis based on DIS jets the experimental uncertainty of $\alps$ is defined by that change in $\alps$ which increases the minimal $\chi^2$ by one unit. The strong coupling is extracted individually from the inclusive jets at low $Q^2$ and from the inclusive, 2-jet and 3-jet at high $Q^2$. The experimentally most precise determination of $\alpha_s(M_Z)$ is derived from the combined fit to all three observables at high $Q^2$. The extracted value is slightly lower than that obtained from photoproduction jets by ZEUS, but compatible within two standard deviations:

\begin{eqnarray}
\nonumber
\alpha_s(M_Z) = 0.1168 ~\pm 0.0007 \,\mathrm{(exp.)}
~ ^{+0.0046}_{-0.0030}\,\mathrm{(th.)}~ \pm 0.0016\,(\textnormal{\scshape pdf})\,. 
\end{eqnarray}
\noindent

 The theory uncertainty is estimated by the \textit{offset method} adding in quadrature the deviations due to various choices of scales and hadronisation corrections. The largest contribution was the theoretical uncertainty arising from terms beyond NLO which amounts to 3$\%$. The PDF uncertainty, estimated using CTEQ6.5, amounts to $1.5\%$. The $\chi^2$ \textit{variation method} leads to smaller uncertainties estimate than the \textit{offset method} for the experiment, while the \textit{offset method} leads to more conservative uncertainties estimate than the \textit{Jones et al.} method for the theory. 
 
\begin{figure}

\psfig{figure=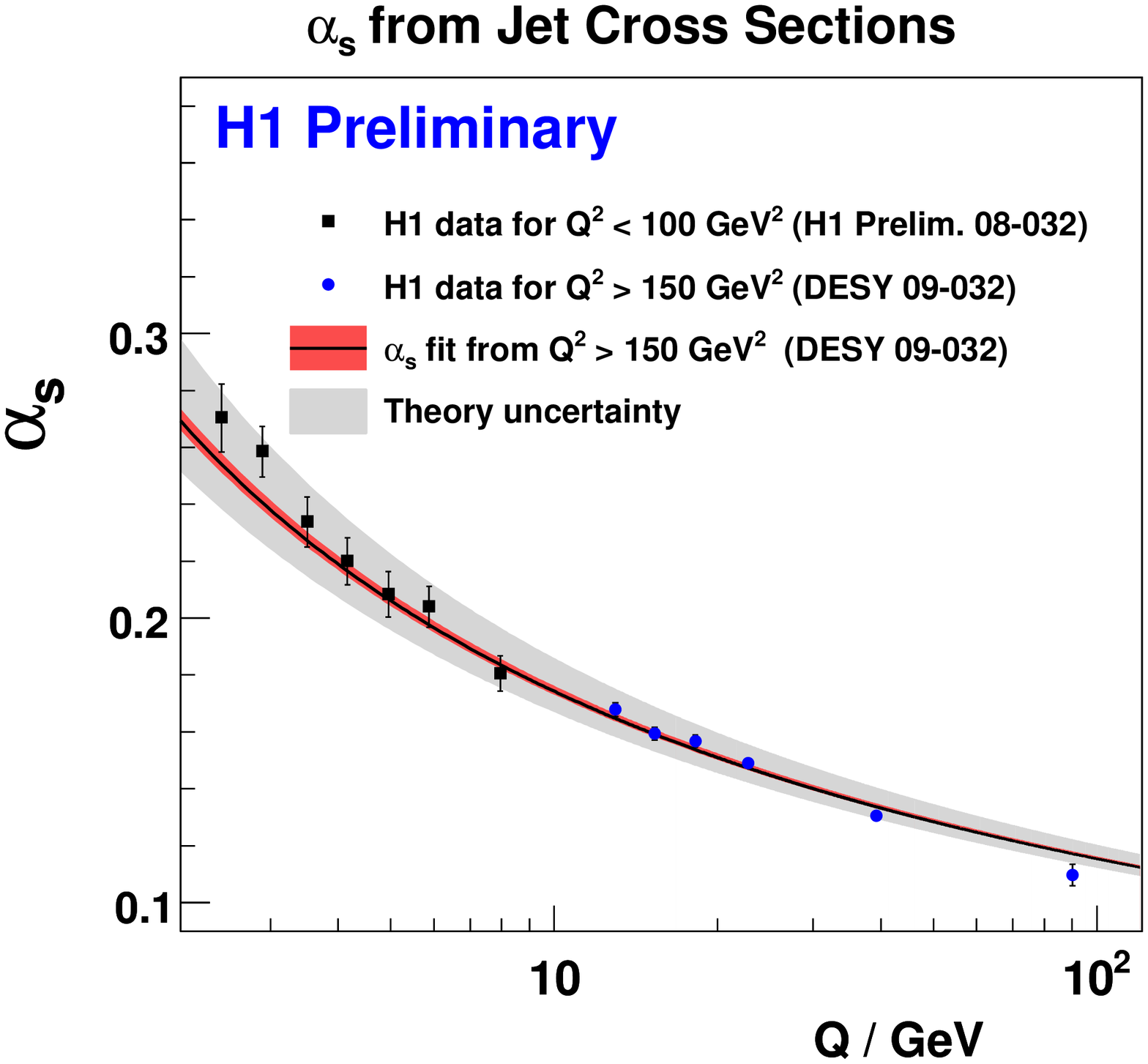,height=2.2in}
\hskip 1.cm
\psfig{figure=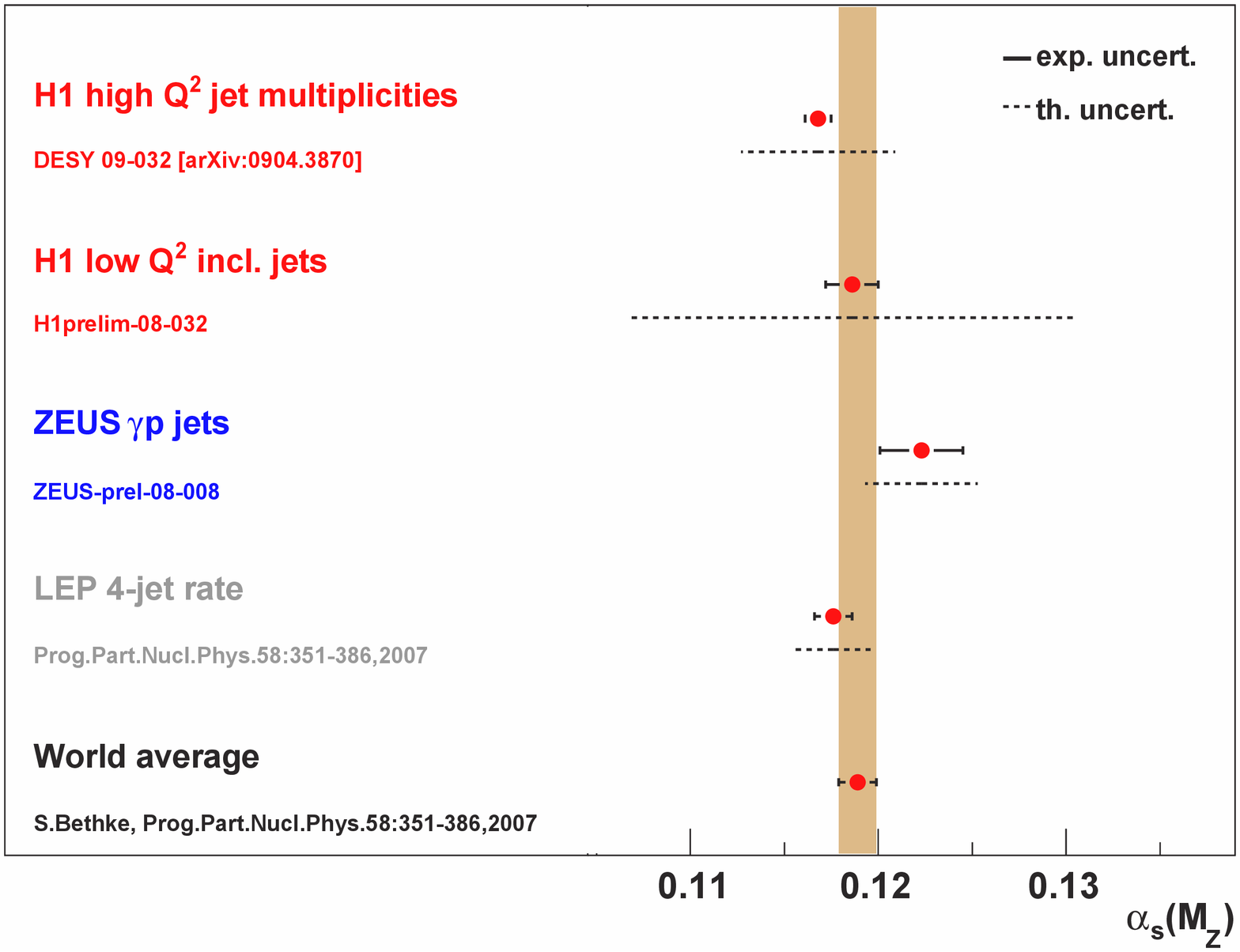,height=2.3in}
\begin{flushleft}
\caption{The running of $\alpha_s(Q)$ (left) and different recent extractions of $\alpha_s(M_Z)$ from HERA, compared to a LEP measurement and the world average.
\label{fig:alphas}}
\end{flushleft}
\end{figure}

 The value extracted at low $Q^2$, \mbox{$\alpha_s(M_Z) = 0.1186 ~\pm 0.0014 \,\mathrm{(exp.)}
 ^{+0.0132}_{-0.0101}\,\mathrm{(th.)} \pm 0.0021\,(\textnormal{\scshape pdf})$}, is compatible with high $Q^2$, but the uncertainty arising from the renormalisation scales variation reach 10$\%$.  The measurement of the strong coupling in a large $Q^2$ range allows to test the $\alpha_s(Q)$ running between 2 and 100 GeV as shown on the figure \ref{fig:alphas}.

	The results for jets at HERA, summarised in figure \ref{fig:alphas}, are competitive with those from $e^+ e^-$ data~\cite{bib:BETHKE} and are in good agreement with different world averages~\cite{bib:BETHKE,bib:WORLD}.

\end{document}